# Planning The Electron Traffic In Semiconductor Networks: A Mesoscopic Analog Of The Braess Paradox Encountered In Road Networks

S. Huant[1], S. Baltazar[2,5], P. Liu[1], H. Sellier[1], B. Hackens[3], F. Martins[3], V. Bayot[1,3], X. Wallart[4], L. Desplanque[4], M.G. Pala[2]

[1]*Institut Néel, CNRS & Université Joseph Fourier, BP 166, F-38042 Grenoble, France*
[2]*IMEP-LAHC, Grenoble INP, Minatec, BP 257, F-38016 Grenoble, France*
[3]*Université Catholique de Louvain, IMCN/NAPS, Louvain-la-Neuve, Belgium*
[4]*IEMN, UMR CNRS 8520, UST Lille, BP 60069, F-59652 Villeneuve d'Ascq, France*
[5]*Departamento de Fısica, Universidad de Santiago de Chile, Santiago 9170124, Chile*

**Abstract.** By combining quantum simulations of electron transport and scanning-gate microscopy, we have shown that the current transmitted through a semiconductor two-path rectangular network in the ballistic and coherent regimes of transport can be paradoxically degraded by adding a third path to the network. This is analogous to the Braess paradox occurring in classical networks. Simulations reported here enlighten the role played by congestion in the network.

## INTRODUCTION

Adding a new road to a congested network can paradoxically lead to a deterioration of the overall traffic situation, i.e. longer trip times for road users. Or, in reverse, blocking certain streets in a complex road network can surprisingly reduce trip time.[1] This counter-intuitive behavior has been known as the Braess paradox.[2] Later extended to networks in classical physics such as electrical or mechanical networks[3], this paradox lies in the fact that adding extra capacity to a congested network can counter-intuitively degrade its overall performance.

Known so far in classical networks only, we have recently extended the concept of the Braess paradox to the quantum world.[4] By combining quantum simulations of a model network and scanning-gate experiments[5-9], we have discovered that an analog of the Braess paradox can occur in mesoscopic semiconductor networks, where electron transport is governed by quantum mechanics.

## EVIDENCE FOR A BRAESS PARADOX IN SEMICONDUCTOR NETWORKS

We have set up a simple two-path network in the form of a rectangular corral connected to source and drain via two openings (see Fig. 1(a) for the network topology).[4] In practice, this corral was patterned from a GaInAs-based heterostructure. The dimensions were chosen to ensure that the embedded 2DEG is in the ballistic and coherent regimes of transport at 4.2K. The short wires in the corral were chosen to be narrower than the source/drain openings to behave as congested constrictions for propagating electrons. We have branched out this basic network by patterning a central wire (see Fig. 1(a)). This opens a third path to the electrons that bypasses the antidot of the initial corral. Then, we have used scanning-gate microscopy to partially block by gate effects the transport through the additional branch. Doing so should intuitively result in a decreased current transmitted through the device, but we just found the opposite behavior in certain conditions, both experimentally and in quantum simulations.[4] Therefore, in a naive picture, electrons in such networks turn out to behave like drivers in congested cities: blocking one path favors their "traffic".

This first finding is summarized in Figs. 1(a) and 1(d), which show the network geometry and a calculated conductance crosscut as function of tip position, respectively. Here, the geometrical parameters are slightly different from those of Ref. 4, namely the outer width and length of the initial corral are 0.75 and 1.6 μm, respectively, whereas the widths of the lateral, upper/lower, and central (additional) arms are $W$=140 nm, $L$=180 nm, and $W_3$=160 nm, respectively. The width of the source and drain openings are $W_0$=320 nm large. This ensures that electron flow in the lateral arms (in the absence of the central arm) is congested because $2W < W_0$. In other words, all injected conduction channels (about 10) into the network cannot be admitted in these arms.[4]

The crosscut in Fig. 1(d) is obtained by computing the network conductance (source-drain voltage= 1 mV) as function of the tip position scanned along the median line of Fig. 1(a). This line crosses the lateral and central arms. Like in Ref. 4, the tip potential is mimicked by a point-like potential of -1V placed at 100 nm above the 2DEG, which corresponds to a lateral extension of $\approx 400$ nm for the tip-induced potential perturbation at the 2DEG level.[4] This model potential entirely depletes the 2DEG in one arm when the tip passes above it.

It is clear in Fig. 1(d) that depleting the central arm produces a distinctive conductance peak that goes well beyond the unperturbed value. This is just the counter-intuitive Braess-like behavior mentioned above. In turn, closing the lateral arms reduces the conductance, in agreement with the intuitive expectation.

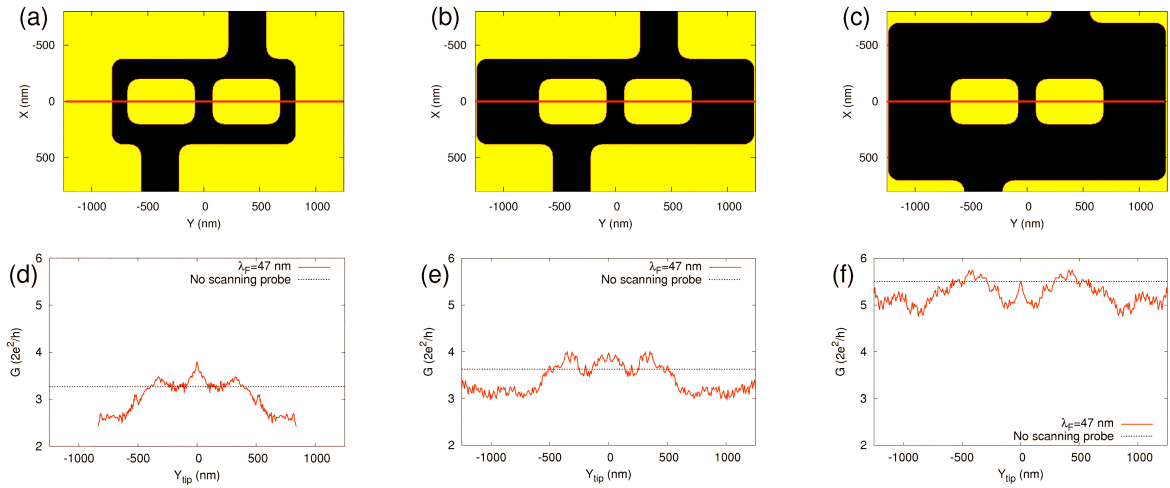

**FIGURE 1.** (a-c) depict the network geometries with parameters given in the text. (d-f) are the corresponding calculated conductance crosscuts in the presence of a depleting tip scanning along the median lines of (a-c). The horizontal dotted lines give the unperturbed conductance without tip. The Fermi wavelength is 47 nm.[4] Fluctuations in the conductance profiles are universal conductance fluctuations (UCFs) due to the tip-induced change in the potential felt by electrons propagating through the device.[10] UCFs are observed for any tip position along the median lines.

## THE ROLE OF CONGESTION

Congestion plays a key role in the occurrence of the classical Braess paradoxes.[1-3] In order to probe a similar role in the mesoscopic counterpart paradox, we have simulated two additional networks with enlarged lateral arms (Figs. 1(b) and 1(e): $W$=560 nm, $L$ unchanged) and with both enlarged lateral and upper-lower arms (Figs. 1(c) and 1(f): $W$=560 nm, $L$=500 nm). This releases congestion in the lateral arms. It is clear from Figs. 1(e) and (f) that releasing congestion smoothens the counter-intuitive conductance peak seen in the congested network when the central arm is blocked. Nevertheless, there is still a slight conductance (current) increase when the tip scans just above the central arms in networks (e) and (f), but this no longer goes beyond the unperturbed conductance for the largest network (f). This finding entails the particular roles played by the additional branch and by network congestion in the occurrence of a distinctive Braess-like paradox. Yet, more experimental and theoretical work is needed to put forward a conclusive explanation at the microscopic level for the paradoxical behavior reported here.

## REFERENCES

1. H. Youn *et al.*, *Phys. Rev. Lett.* **101**, 128701 (2008).
2. D. Braess, *Unternehmensforschung* **12**, 258-268 (1968); D. Braess, A. Nagurney, and T. Wakolbinger, *Transp. Sci.* **39**, 446-450 (2005).
3. C. M. Penchina and L. J. Penchina, *Am. J. Phys.* **71**, 479-482 (2003).
4. M.G. Pala *et al.*, *Phys. Rev. Lett.* **108**, 076802 (2012).
5. M.A. Topinka *et al.*, *Science* **289**, 2323-2316 (2000).
6. A. Pioda *et al.*, *Phys. Rev. Lett.* **93**, 216801 (2004).
7. B. Hackens *et al. Nature Phys.* **2**, 826-830 (2006); H. Sellier *et al.*, *Semicond. Sci. Technol.* **26**, 064008 (2011).
8. F. Martins *et al.*, *Phys. Rev. Lett.* **99**, 136807 (2007); M.G. Pala *et al.*, *Phys. Rev. B* **77**, 125310 (2008).
9. B. Hackens *et al.*, *Nat. Commun.* **1**, 39 (2010).
10. R.A. Webb *et al.*, *Phys. Rev. Lett.* **54**, 2696 (1985).